\documentclass[11pt]{article}
\usepackage{amsmath,amsfonts,amssymb,amsthm,url}
\usepackage{graphicx}
\usepackage{authblk}

\usepackage{tikz}
\usetikzlibrary{shapes,arrows}
\usepackage{pgfplots}

\def\Zahl{\mathbb{Z}}
\def\Cli{\mathcal{C}}
\def\Ser{\mathcal{S}}
\def\Par{\mathcal{P}}
\def\cred{\mathrm{cred}}

\title{Privacy-Preserving Trust Management Mechanisms from Private Matching Schemes}
\author{Oriol Farr\`as}
\author{Josep Domingo-Ferrer}
\author{Alberto Blanco-Justicia}
\affil{Universitat Rovira i Virgili\\ 
Department of Computer Engineering and Maths\\
UNESCO Chair in Data Privacy\\
Av. Pa\"{\i}sos Catalans 26, 43007 Tarragona, Catalonia\\ 
\{oriol.farras,josep.domingo,alberto.blanco\}@urv.cat}

\date{}

\begin{document}
\maketitle

\begin{abstract}
Cryptographic primitives are essential for constructing pri\-va\-cy-pre\-serv\-ing communication mechanisms. There are situations in which two 
parties that do not know each other need to exchange 
sensitive information on the Internet. Trust management mechanisms make use of digital credentials and certificates in order to establish trust among these strangers. We address the problem of choosing which credentials are exchanged. During this process, each party should 
learn no information about the preferences of the other party other than strictly required for trust establishment. 
We present a method to reach an agreement on the credentials to be exchanged that preserves the privacy of the parties. Our method is based on secure two-party computation protocols for set intersection. Namely, it is constructed from private matching schemes.
\end{abstract}


\section{Introduction} 

Interactions between parties that involve exchanging 
sensitive information are part of everyday life. Taking a medical test, paying with a credit card or asking for directions are examples of such interactions. In all of these cases an individual or organization $\Cli$ reveals some information to another individual or organization $\Ser$ so that $\Ser$ can provide a service to $\Cli$. 
Clearly, an exchange of personal information is more likely to take place if there is trust between the interacting parties. 
For instance, people agree on revealing medical data
to a doctor in a medical center, but not 
to anyone or anywhere.
These interactions are easy to carry out face to face and in a specific context,
but they are challenging if performed over the Internet, where
personal identification is not obvious and the physical context
is simply not there.

A first approach is securing the communication using cryptographic protocols. 
Using these techniques in combination with public key infrastructures provides users interacting with remote parties with the certainty that they are communicating with the real service provider. Furthermore, encrypting communication 
prevents third parties from eavesdropping on the transmitted contents.
This has been the basis of secure digital communications and e-commerce, but 
recent reports show that authentication is not always enough for users to trust service providers~\cite{mef,EUBARO}. 


Therefore, there is a need to design new access control systems in which not only the identity of the parties is revealed and assured, but trust is built through the exchange of valid credentials that contain attributes of the parties. Trust management mechanisms make use of digital credentials and certificates in order to establish trust among strangers. Trust negotiation schemes are protocols for establishing trust between parties unknown to each other 
through the exchange of credentials and personal 
information; in
such negotiation protocols, the disclosure of this information is performed according to access control policies determined by the parties.

\subsection{Motivation}

The special Eurobarometer on data protection 
and electronic identity~\cite{EUBARO} shows some interesting results regarding the perception of privacy by citizens of the European Union. Specifically, 
74\% of the Europeans see the disclosure of personal information as a part of modern life. The main reason for disclosure is to access online services, although 43\% of the respondents claim they have been asked for more personal information than necessary in order to access these services. Finally, the report shows that a majority of Europeans are concerned about their behavior being
recorded via payment cards, mobile phones or mobile Internet.

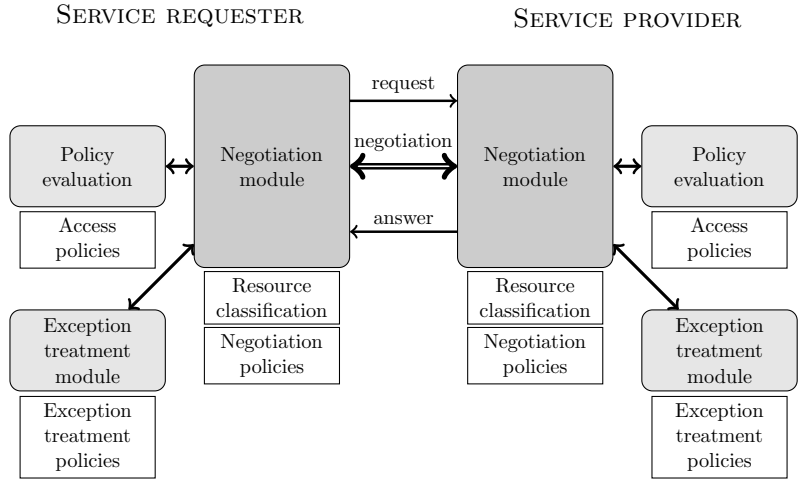
\begin{figure} 
\centering

 \tikzstyle{block} = [rectangle, draw, fill=black!10, 
    text width=7em, text centered, rounded corners, minimum height=4em]
 \tikzstyle{nego} = [rectangle, draw, fill=black!20, 
    text width=7em, text centered, rounded corners, minimum height=10em]
\tikzstyle{line} = [draw, -latex']
\tikzstyle{cloud} = [draw, rectangle,text centered, text width=6em, node distance=2.5cm, minimum height=2em]

\label{neg}
\begin{tikzpicture}[node distance = 3.5cm, scale=0.7, every node/.style={scale=0.7}]
  \node [nego] (neg1) {Negotiation module};
  \node [cloud,below of=neg1](r1){Resource classification};
  \node [cloud,below of=r1,node distance=1.1cm] {Negotiation policies}; 
  \node [block, left of=neg1] (p1) {Policy evaluation};
  \node [cloud,below of=p1,node distance=1.4cm] {Access policies}; 
  \node [block, below of=p1,node distance=3.5 cm](e1){Exception treatment module};
   \node [cloud,below of=e1,node distance=1.65cm] {Exception treatment policies}; 
   
  \node [nego, right of=neg1,node distance=5cm] (neg2) {Negotiation module};
  \node [cloud,below of=neg2](r2){Resource classification};
  \node [cloud,below of=r2,node distance=1.1cm] {Negotiation policies}; 
  \node [block, right of=neg2] (p2) {Policy evaluation};
  \node [cloud,below of=p2,node distance=1.4cm] {Access policies}; 
  \node [block,below of=p2,node distance=3.5 cm](e2){Exception treatment module};
   \node [cloud,below of=e2,node distance=1.65cm] {Exception treatment policies}; 
  
    
  \path [line,very thick,<->](neg1)--node [above=2.5cm] {\Large{\textsc{Service requester}}}(p1);
  \path [line,very thick,<->](neg1)--(e1);
  \path [line,very thick,<->](neg2)--node [above=2.5cm] {\Large{\textsc{Service provider}}}(p2);
  \path [line,very thick,<->](neg2)--(e2);
    \path [line,->,line width=0.2ex](neg1.40)-- node [above] {request} (neg2.140) ;
  \path [line,double,line width=0.2ex,<->](neg1.00)-- node [above=4pt,node distance=1.1cm] {negotiation} (neg2.180) ;
    \path [line,line width=0.2ex,<-](neg1.320)-- node [above] {answer} (neg2.220) ;
    
\end{tikzpicture}
\caption{Interoperable Trust Assurance Infrastructure}
\label{intertrust}
\end{figure}

Trust management is treated as a building block of many commercial frameworks.
One example is the Interoperable Trust Assurance Infrastructure 
(Inter-Trust)~\cite{it}. It is a project that seeks to develop a framework to support trustworthy applications in heterogeneous networks and devices based on the enforcement of interoperable and changing security policies (Figure~\ref{intertrust}). In Inter-Trust, the trust negotiation is essential to reach agreements on the security policies,
the so-called Service Level Agreements. Inter-Trust will incorporate trustworthiness by integrating legal, social and economic concerns, allowing applications and devices to negotiate and be constrained by them.


\subsection{Privacy-Preserving Trust Management}

A critical issue in trust management is to preserve the privacy of the users.
During a trust management process, the parties can try to learn information about each other.
On the one hand, the parties can try to obtain information about the preferences of the service providers. The requester of a service can ask for different options to access the service. Then the server must provide the different acceptable options. Since the revealed options may reflect the business model and the target customers considered by the provider, service providers are reluctant to show a full description of their access policy. 

On the other hand, requesters do not want to provide information on the credentials they own unless those credentials are essential for the transaction. 
Clients are willing to show a credential if needed for accessing a service, otherwise
they will withhold it in order to preserve their privacy. 
Further, each requester has a specific access control policy for each credential. That is, a requester will show a credential only if the service provider gives enough evidences of fairness.

In summary, service providers are reluctant to show their access policies, and clients want to disclose as little private information as possible. 
Therefore, during the trust management process each party should learn 
no information about the access policies or preferences of the other parties 
beyond what is strictly required for trust establishment.
Solutions based on trust negotiation mechanisms~\cite{DD,FLA,LWBW,LWP,LLW,NOW,SBFPT,WSJ} control the disclosure of user preferences by showing the access control policies in a sequential way. However, trust negotiation mechanisms are oriented to controlling credential disclosure, and the users may obtain information on the access control policies by 
playing with the system. Trust management mechanisms based on secure multiparty computation~\cite{MWI,NBMW,Yao06} provide higher privacy protection.

\subsection{Our Results}

We address the problem of constructing a privacy-preserving mechanism for choosing the credentials to be exchanged.  Moreover, we consider that privacy preservation should be achieved as effortlessly as possible. Therefore, our goal is to 
come up with an efficient and privacy-preserving mechanism to determine
the optimal set of informations to be disclosed, 
according to the preferences of the two parties. The exchange of credentials and the final applications are beyond the scope of this work, 
which focuses on trust management.

We present a method to reach an agreement on the credentials to be exchanged that preserves the privacy of the parties. We consider the two-party case,
in which one party is a client that wants to access a service, and the other party is the server that provides it. Our method is based on secure two-party computation protocols for set intersection. Specifically, it is constructed from the private matching schemes in~\cite{FNP}. 

The client sends a list of options to the server in a private way.
Each option is a combination of credentials the client would agree to show.
The server has a correspondence list that, for each accepted combination of client credentials, specifies the credentials the server would show. 
Using secure multiparty computation techniques, client and server 
compute the matching options. Then the server sends to the client the options that match 
the client's preferences. In this way, the server does not learn the 
preferences of the client, and the client only learns the specific access 
policies that match her selected options. 

The secure two-party computation primitive we use is based on homomorphic encryption. 
Using the Paillier cryptosystem~\cite{Pai}, the total number of exponentiations needed is $O(s+t\ln\ln s)$, where $s$ and $t$ are the number of options 
specified by the client and the server, respectively.

The rest of this paper is organized as follows. In Section~\ref{ATN} we present an introduction to trust management. Section~\ref{SMP} is devoted to secure multiparty computation and to private matching schemes. We present our method in Section~\ref{OR}. 
Section~\ref{RW} lists conclusions and open problems. 

\section{Trust Management}
\label{ATN}

Remote communications over the Internet often require the interacting parties to trust each other, especially when the communication involves the exchange of private, confidential, or sensitive information. Traditional approaches to establish trust assume that the parties are known to each other before the communication takes place. Organizations often sign a Service Level Agreement (SLA) and collaboration contracts before engaging in the exchange of services and information. This approach is not always possible, because the assumption that the parties are known to each other is not always true, especially in open environments
such as the Internet and the Web~\cite{WSJ}.

Secure Sockets Layer (SSL) and its successor Transport Layer Security (TLS) 
are cryptographic protocols that provide trust and security to communications over the Internet. These protocols begin with a negotiation or {\em handshake} phase in which the two parties (normally a client and a server) agree on an
encryption algorithm and a shared key to encrypt the communication. Also, during this phase the two parties exchange digital certificates in order to authenticate each other. Even though the use of TLS is widespread, users do not fully trust Internet service providers, as discussed in the previous section. 
Therefore, there is a need to improve existing strategies and/or 
devise new methods for establishing trust.

More recent approaches to establishing trust are the {\em Automatic Trust Negotiation} (ATN) protocols. ATN is based on the exchange of digitally signed credentials to establish trust and make access control decisions. Digital credentials are an extension of traditional electronic certificates that only prove the identity of a user. Credentials can include additional attributes, and hence 
they can certify more properties of that user, such as the age, the permission to perform a certain activity, membership to a certain organization, etc. These credentials can be seen as an electronic version of the physical credentials that everyone carries in their wallets~\cite{Win}. 

Although introducing such additional attributes can play an important role in the establishment of trust, it is important to note that they can represent sensitive information by themselves, and should only be shared with other parties according to some access control policies. For example, a user $\Cli$ might require some credentials from $\Ser$ that certify that $\Ser$ is a doctor in order to share her medical history. In case $\Ser$ cannot deliver the required credentials to the user $\Cli$, the exchange of credentials will be interrupted 
and the negotiation will fail.

Several tools and techniques have been presented to avoid such deadlock situations, for instance~\cite{LLW,OACert,HBS,TAC}. These tools can be classified
into cryptographic credentials, access control policies and 
negotiation techniques. 
The rest of this section describes these three tool classes 
in more detail.

\subsection{Electronic Credentials}

As pointed out in~\cite{Win}, digital versions of the same credentials we carry in our pockets should be enough to devise similar trust mechanisms over the Internet as we normally do in everyday situations. Using Public Key Infrastructures (PKI) and Trusted Third Parties (TTP), it is possible to issue digital credentials that can be verified in a secure way.

With this in mind, X.509 certificates come as a reasonable tool to prove identity and other properties about the user~\cite{X509}, and as such, they have become the standard for authentication in the Internet. However, X.509 was not designed with privacy in mind. Li et al.~\cite{LLW} enumerated the properties a digital certificate should offer and the related cryptographic tools for 
carrying out trust negotiations while preserving the privacy of the parties.
These properties are: separation of credential and attribute disclosure, selective disclosure of attributes, oblivious use of credentials and attributes, and zero-knowledge proofs and secure multiparty computation schemes to compute and prove compliance with policies.

Oblivious Attribute Certificates (OACerts, \cite{OACert}) are a good example of a digital credential scheme that complies with these properties. OACerts allow parties to use attribute values stored in the certificate in an oblivious way. The attributes can be used to evaluate Boolean conditions in policies without revealing the actual values. The oblivious use of credentials is also studied in~\cite{FAL,FLA,HBS,LLW}.

\subsection{Access Control Policies}

Policies control the disclosure of attributes and credentials of the parties. These policies range from the simple case of identification of the requester to more complex systems that involve users, roles and organizations with different levels of permissions towards the offered resources.

Access control policies have been a topic of discussion in both industry and academia. W3C and OASIS both have published a set of standards that regulate several types of policies to control the access to and the use of Web services. Of these the most popular one is probably the XML-based XACML~\cite{XACML}.

Other important examples of access control models are attribute-based access control (ABAC), of which XACML is an example, discretional access control (DAC), mandatory access control (MAC), role-based access control (RBAC), organizational role-based access control (OrBAC)~\cite{OrBAC}, etc.

\subsection{Automated Trust Negotiation Systems}

By adding some additional tools and mechanisms to the previously mentioned
ones, a number of final systems have been developed. In this 
section we present a few examples

TrustBuilder~\cite{LWP} is focused on trust negotiation and uses an ABAC model. The TrustBuilder approach consists in sequentially disclosing credentials and policies in a controlled way, such that sensitive credentials and policies are only disclosed once a certain level of trust is established. Policies are only disclosed if they are relevant to the negotiation.

The Trust$\mathcal{X}$ framework~\cite{SBFPT} makes use of an XML-based language to define credentials and policies. Parties using Trust$\mathcal{X}$ collectively build a tree that represents all the possible ways in which 
the negotiation can take place. A path from the root to a leaf that satisfies both parties is the accepted negotiation process.

XeNA~\cite{HCCD} is based on XACML and RBAC. The system sequentially exchanges credentials to fulfill the conditions of the access control policies. Credentials are classified by levels or {\em classes} of sensitivity.
 MoTOrBAC~2~\cite{ACCC} is a security administration tool. It allows the specification and administration of OrBAC-based security policies.
Finally, Idemix~\cite{CH}, allows creating and managing anonymous credentials.
%

\section{Secure Multiparty Computation}
\label{SMP}

Secure multiparty computation allows a set of parties to compute a
joint function of their inputs in a secure way without requiring a trusted 
third party. During the execution of the protocol the parties do 
not learn anything about each other's input except what is implied by the 
output itself. 
There are two main adversarial models: honest-but-curious adversaries and malicious adversaries. In the former model, the players follow the protocol's instructions but try to obtain information about other players' inputs from the messages they receive. In the latter model, 
the adversary may deviate from the protocol in an arbitrary way. 
Aumann and Lindell~\cite{AL} introduced a new model, the covert adversaries. A covert adversary may deviate from
the protocol in an attempt to cheat, but such deviations are detected by honest parties. In this context, the parties may be considered rational, that is,
acting according to their interests. 
In game-theoretic terms, it is assumed that players only try to maximize their utility functions; hence, all possible deviations from the correct protocol
execution have this goal.


This work is restricted to the two-party case and the setting in which
both parties have the same output.
As to the type of computations to be carried out,
we only consider the set intersection function and 
other combinatorial functions. For convenience, we will distinguish 
the role of the two parties: one will be a 
client $\Cli$ and the other a server $\Ser$.


The intersection of two sets can be obtained by using the generic constructions based on Yao's garbled circuit~\cite{Yao86}. This technique is very generic, because it
allows computing any arithmetic function, but for most of the functions it is inefficient. Many of the recent works on two-party computation are focused on improving the efficiency of these protocols for particular families of functions.

Freedman, Nissim, and Pinkas~\cite{FNP} presented a more efficient method to compute the set intersection that uses polynomial-based techniques and homomorphic encryption. Specifically, they presented a {\em private matching scheme}, 
a two-party protocol that works as follows. First 
the client $\Cli$ and the server $\Ser$ agree on a finite domain of 
elements $D$. Then
$\Cli$ inputs $X=\{a_1,\ldots,a_s\}\subseteq D$,
$\Ser$ inputs $Y=\{b_1,\ldots,b_t\}\subseteq D$, 
where $s$ and $t$ are known, and finally
$\Cli$ learns $X\cap Y$.  
The scheme can be constructed by means of the Paillier cryptosystem~\cite{Pai}.
It exploits the property that given three elements $m_1, m_2, m_3$,
it is possible to compute efficiently $Enc(m_1+m_2)$ and $Enc(m_1\cdot m_3)$ from $Enc(m_1)$,  $Enc(m_2)$, and $m_3$. 

Next we present an outline of the scheme in~\cite{FNP} that is secure against
in the honest-but-curious model.
\begin{enumerate}
\item $\Cli$ computes the polynomial $p(x)=\prod_{i=1}^s (x-a_i)$. 
\item $\Cli$ sends $Enc(p_0),\ldots, Enc(p_s)$ to $\Ser$, where $p_i$ is the coefficient of degree $i$ of $p$.
\item For every $1\leq j\leq t$, $\Ser$ picks a random element $r_j\in\Zahl_n$
and computes $Enc(r_j\cdot p(b_j)+b_j)$. Observe that these ciphertexts can be obtained from $Enc(p_0),\ldots, Enc(p_s)$ and $Enc(b_j)$ by using the homomorphic properties of the cryptosystem.
Then $\Ser$ sends the ciphertexts to $\Cli$.
\item $\Cli$ decrypts the $t$ ciphertexts. The result of each decryption is an 
element from $X\cap Y$ or a random element.
\end{enumerate}
If  the size of the domain of $Enc$ is much larger than $|X|$
the scheme computes $X\cap Y$ with high probability.
There is a variant of this protocol that allows adding an additional 
payload. Step 3 of the 
protocol described above is replaced by the following one:
\begin{enumerate}
\item[3'.] $\Ser$ picks a random element $r_j\in\Zahl_n$ for every $1\leq j\leq t$. 
Let $c_j$ be a payload data associated to $b_j$ that $\Ser$ wants to send to $\Cli$. 
Then $\Ser$ computes $Enc(r_j\cdot p(b_j)+(b_j||c_j))$ for $1\leq i\leq t$, and
sends these ciphertexts to $\Cli$.
\end{enumerate}
Observe that in this case, in the last step $\Cli$ decrypts the $t$ ciphertexts 
and obtains $X\cap Y$ and an additional information for each of these elements.

The idea of  Freedman, Nissim, and Pinkas~\cite{FNP} was used in many other works to improve the computation of set operations. Kissner and Song~\cite{KS} presented secure multiparty computation protocols for computing the set 
intersection, multi-set intersection and other combinatorial operations.  They present constructions for honest-but-curious adversaries and malicious adversaries. 
Hazay and Lindell~\cite{HL} presented a construction that is secure in the 
covert model. There are also other interesting constructions in the 
literature, such as~\cite{DMRY,MR}.



\section{A Privacy-Preserving Trust Management Scheme}
\label{OR}

In this section we present a new mechanism for privacy-preserving 
trust management. The situation we consider is the following one. A client $\Cli$
wants to buy a service from a server $\Ser$. $\Ser$ needs 
some personal and financial information about $\Cli$ to perform the transaction.
However, $\Cli$ is reluctant to show private information to $\Ser$, 
because $\Cli$ is not sure that $\Ser$ trustworthy.

The mechanism we construct is a protocol based on the private 
matching scheme presented in the previous section. 
Namely, the protocol with the alternate step $3'$. Hence 
our protocol is secure in the honest-but-curious model. 
For the sake of completeness, we present herein all the
details of our proposal.

Our proposal allows parties $\Cli$ and $\Ser$ 
to agree on the information they have to exchange to perform the
transaction in a private way. Broadly speaking, 
$\Cli$ first sends an encrypted message
to $\Ser$ that declares which credentials 
and personal information she would be inclined to reveal to $\Ser$.
$\Ser$ cannot read the message, but he can create an 
encrypted message containing the options declared
by $\Cli$ in which he agrees, and the information $\Ser$
would reveal in each case. The interest of 
the protocol lies in the protection of the preferences
of each party. 
That is, $\Ser$ does not learn the 
preferences of $\Cli$, and $\Cli$ only learns the specific access 
policies that match her selected options. 

Let $E_{\Cli}$ and $E_{\Ser}$ be the domains of 
credentials and personal data of $\Cli$ and $\Ser$, 
respectively. Define $D_{\Cli}=\Par(E_{\Cli})$ and
$D_{\Ser}=\Par(E_{\Ser})$. Notice that, for any set $A$, 
$\Par(A)$ is the family of subsets of $A$.

First $\Cli$ defines different combinations of 
elements from $E_{\Cli}$ that she would be ready to show to $\Ser$.
Let $X=\{a_1,\ldots,a_s\}\subseteq D_{\Cli}$ be the set of 
such options. Independently, $\Ser$ 
defines $Y=\{(b_1,c_1),\ldots,(b_t,c_t)\}\subseteq D_{\Cli}\times D_{\Ser}$,
the acceptable combinations $(b_i,c_i)\in D_{\Cli}\times D_{\Ser}$ 
according to his preferences. 
That is, for every acceptable combination of elements $b_i$ from $D_{\Cli}$, 
$\Ser$ would show $c_i\in D_{\Ser}$. Observe that 
$(b_i,c_i)\neq (b_j,c_j)$ for every $1\leq i<j\leq s$, but $b_i$ and $b_j$
(or $c_i$ and $c_j$) may be equal.
Now we present our protocol.

\begin{enumerate}
\item $\Cli$ computes the polynomial $p(x)=\prod_{i=1}^s (x-a_i)$. 
\item $\Cli$ sends $Enc(p_0),\ldots, Enc(p_s)$ to $\Ser$, where $p_i$ is the coefficient of degree $i$ of $p$.
\item For every $1\leq j\leq t$, $\Ser$ picks a random element $r_j\in\Zahl_n$
and computes $Enc(r_j\cdot p(b_j)+(b_j||c_j))$. Then $\Ser$ sends the ciphertexts to 
$\Cli$.
\item $\Cli$ decrypts the $t$ ciphertexts. 
\end{enumerate}
The result of each decryption is an 
element from $X$ attached to an element of $D_{\Ser}$ 
or a random element.
 
\subsection{Discussion}
 
As encryption method we can use the Paillier cryptosystem~\cite{Pai}. In this case, we take $n=p\cdot q$, where $p$ and $q$ are large primes satisfying the properties  in~\cite{Pai}. Then we describe $X\in\Zahl_n$ and $Y\subseteq \Zahl_n\times \Zahl_n$. A way to encode an option in $\Zahl_n$ is the following. First, we establish an order among the credentials. Given an option $\{\cred_{i_1},\cred_{i_2},\ldots,\cred_{i_u}\}$ for some $i_1<i_2<\ldots<i_u$,
we consider $x=\sum_{j=1}^u 2^{i_j}$. If the domain $D_{\Cli}$ (or $D_{\Ser}$) is much larger than the number of realistic options, we can use a hash function~\cite{FNP}. The amount of exponentiations needed is
$O(s\cdot t)$, and it can be reduced to $O(s+t\ln\ln s)$~\cite{FNP}. 

The protocol is secure in the honest-but-curious adversary model. Following~\cite{FNP} we can create a protocol that is secure in the malicious adversary model
by means of zero-knowledge proofs. The resulting protocol is much less efficient.

We consider that the previous solution is tailored according to the common
client-server context in which the client is usually at a disadvantage. It grants a higher protection to the privacy of the client than to the privacy of the server.  However, there are other situations in which we need to guarantee
a more equitable treatment. In this case, private matchings also provide a natural
solution for privacy-preserving trust management. A solution would be 
a private matching in which the inputs contain the preferred options about 
one's own credentials and the other party's credentials. That is, $X,Y\subseteq D_{\Cli}\times D_{\Ser}$. A solution along
this line was presented in~\cite{MWI}. 

In this work we present a method for agreeing on the credentials to be exchanged, 
but we do not analyze the way the credentials are exchanged and disclosed.
There are many schemes for {\em fair exchange} of information 
between different parties. Some recent proposals consider
schemes that are secure in the covert model, as for instance~\cite{BH,D10,D11}.
Most of these solutions model the exchange of information as a static 
game with complete information, and hence look for Nash equilibria.


%


\section{Related Work}
\label{RW}

Secure multiparty computation protocols have been used as building
blocks to obtain privacy-preserving data mining techniques.
Specifically, they can be used for finding correlations
and patterns among different attributes from large relational databases.
The protocols for set intersection are specially interesting in
data mining. For example, they allow two companies to discover the amount of
customers they share without having to reveal the entire lists of their
customers.

Yao et al.~\cite{Yao06} presented Point-Based Trust, a trust management mechanism built from a tailored secure multiparty computation protocol. The owner of a resource values each credential with an access threshold, a certain number of points. 
This access threshold is the minimum amount of points required 
by the owner from the client to give the latter 
access the resource. The resource owner also defines a point value of each credential, which denotes the number of points a client obtains if she
discloses that credential. The thresholds and the point values are not revealed
by the owner to the client. 
The output of the protocol provides a combination of the credentials that allows the client to obtain access while minimizing her privacy loss. This protocol is secure in the 
honest-but-curious adversarial model
and uses secure two-party protocols for computing the maximum of two values~\cite{FA}. 
A drawback of this scheme is that the quantitative approach is not always the most suitable. For instance, the national id, the passport, and the driving license are documents that provide a similar amount of trust and may have a high weight. However, observe that the amount of trust obtained by disclosing these credentials is not the sum of individual point values, because there is a lot of redundant information between them. Moreover, there are some informations that make sense only if they are presented at the same time. For instance, the expiration date of a credit card is useless without the credit card number.

In a privacy-reconciliation protocol~\cite{MWI,NBMW}, each party holds a private input set in which the elements are ordered according to the party's preferences. The goal of a reconciliation protocol on these ordered sets is to find all common elements in the parties' input sets that maximize the joint preferences of the parties. The main drawback of these schemes is the efficiency. The computation of the best option is, in general, a hard problem and so the protocols are less efficient than the scheme presented here. Adding privacy protection to reconciliation protocols can increase by two orders of magnitude the running time of the reconciliation protocols~\cite{MWI,VIWM}.

\section{Conclusions}

In this paper we have presented a privacy-preserving mechanism for trust management.
This work is restricted to the two-party case. Given the preferences of
each party on credential disclosure, our method provides a proposal on 
the credentials to be exchanged that is consistent with 
the parties' preferences. The privacy of the parties 
is preserved because their preferences are protected 
by a protocol for secure multiparty computation for set intersection that is secure in the honest-but-curious model. 

Future work might consider the combination of this trust management method with fair exchange mechanisms and the integration of these building blocks into more general frameworks. Moreover, it would be interesting to extend this construction to the covert adversarial model. That is, to find a protocol secure in the rational model computing the same function. 


\section*{Acknowledgments}

This work was partly supported by
the Government of Catalonia under grant 2009 SGR 1135, by
the Spanish Government
through projects 
TIN2011-27076-C03-01 ``CO-PRIVACY''
and CONSOLIDER INGENIO 2010 CSD2007-00004 ``ARES'',
and by the European Comission under FP7 project
`Inter-Trust''.
The second author
is partially supported as an ICREA Acad\`emia researcher
by the Government of Catalonia;
he is with the UNESCO Chair in Data Privacy,
but he is solely responsible for the views
expressed in this paper, which do not necessarily
reflect the position of UNESCO nor commit that organization.

\end{document}